\documentclass[prb,floatfix,showpacs,superscriptaddress,longbibliography,twocolumn]{revtex4-1}
\usepackage[USenglish]{babel}
\usepackage{amsmath,amssymb,amsfonts}
\usepackage{graphicx}
\usepackage{subfigure}
\usepackage{import}
\usepackage{color}
\usepackage{epsfig}

\usepackage[colorlinks=true,linkcolor=blue,citecolor=red]{hyperref}

\newcommand{\eq}[2]
{
  \begin{equation}
    #1
    \label{#2}
  \end{equation}
}

\newcommand{\equ}[1]
{Eq.~(\ref{#1})}

\newcommand{\figu}[1]
{Fig.\ref{#1}}

\newcommand{\secu}[1]
{Sec.~\ref{#1}}

\def\bcen{\begin{center}}
\def\ecen{\end{center}}

\def\a{\alpha}              
\def\e{\varepsilon}          
              
                    \def\s{\sigma}

\def\FF{{\cal F}}\def\HH{{\cal H}}

\def\GG{{\cal G}}

  \def\ie{\mbox{\it i.e.\ }}
\def\=={\equiv}

\def\qed{\raise1pt\hbox{\vrule height5pt width5pt depth0pt}}

\def\cG0{{\cal G}_0} 
\def\cG{{\cal G}}

\def\up{\uparrow}  \def\dw{\downarrow}
\def\bra{\langle} \def\ket{\rangle}

  \def\Im{\mbox{Im}}
\def\ie{\hbox{\it i.e.\ }} 
\def\ie{\mbox{\it i.e.\ }} \def\=={\equiv}
 
\def\Im{{\rm Im}}  
 \def\ep0{\epsilon_{p}} \def\ed0{\epsilon_{d}}

\begin{document}
\title{Dynamical Mean-Field Theory description of the voltage induced
  transition in a non-equilibrium superconductor}

\author{A.~Amaricci} 
\affiliation{Democritos National Simulation Center, 
Consiglio Nazionale delle Ricerche, 
Istituto Officina dei Materiali and 
Scuola Internazionale Superiore di Studi Avanzati, 
Via Bonomea 265, 34136 Trieste, Italy}

\author{M.~Capone}
\affiliation{Democritos National Simulation Center, 
Consiglio Nazionale delle Ricerche, 
Istituto Officina dei Materiali and 
Scuola Internazionale Superiore di Studi Avanzati, 
Via Bonomea 265, 34136 Trieste, Italy}

\date{\today}

\begin{abstract}
Using dynamical mean-field theory (DMFT) we study a simplified model for
heterostructures involving superconductors. The
system is driven out-of-equilibrium by a voltage bias, imposed as an
imbalance of chemical potential at the interface. 
We solve the self-consistent DMFT equations using iterative second-order perturbation
theory in the Nambu-Keldysh formalism.  We show that the superconducting
state is destabilized by voltage biases of the order of the energy gap.
We demonstrate that the transition to the normal state occurs through
an intermediate {\it bad superconducting} state, which is
characterized by a smaller value of the order parameter and incoherent
excitations. We discuss the energetic balance behind the stabilization
of such exotic superconducting state.
\end{abstract}


\maketitle

\section{Introduction.}
The unstoppable advances in the engineering of heterostructures based on
transition-metal oxides is setting the pace for their theoretical
investigation. A variety
of phenomena has been reported through the combination of different
geometries and the rich physics of the constituents, which range from
colossal magnetoresistance manganites to high-temperature
superconducting cuprates
\cite{Luo2008APL,Mannhart2007S,Ramesh2010PRL,Sangiovanni2013PRL}. 
There is no need to stress the importance, also for applications
in electronics,  of superconducting
heterostructures and of the possibility to control their conduction
properties via a gate voltage\cite{Schlom2010S,DevoretS2013,Yu2014NP}.
However, even if our ability to theoretically describe strongly correlated
electron systems in confined geometries is reaching remarkable
levels\cite{Sulpizio2014AROMR}, a full understanding of the
non-equilibrium properties of driven
superconducting heterostructures remains a formidable task, which requires
to take some intermediate steps before a full microscopic modelling
can be achieved.

In this work we perform one step in this direction, considering the
ballistic transport in an idealized superconductor subject to an
external bias. 
Extending to the superconducting state a proposal introduced in
Ref.~\onlinecite{Han2009PRB}, we consider a simplified model for the
bulk properties of a system subject to a finite voltage bias. 
The potential drop occurs essentially at the interfaces of the superconductor with the
normal source/drain leads\cite{Keizer2006PRL,Klapwijk2012PRB,Yu2014NP}, which in
turn are located sufficiently far in space, while the potential
profile is essentially constant in the bulk.  
The underlying idea is that the charge imbalance taking place at the interfaces
injects into the system evanescent electronic states \cite{Zenia2009PRL,Borghi2009PRL}
at the bias energy, which rapidly reach a ballistic regime. 
Thus in the stationary limit and for the bulk, the presence of a voltage bias is accounted for through 
the existence of electrons propagating in the same or in the opposite
direction with respect to the driving field, respectively {\it Left}
or {\it Right} movers, with unbalanced energies set by the voltage
bias at the two leads.

Although this simplified construction can not capture the full non-equilibrium physics 
of a driven bulk superconductor, it allows us to get insight
into the effect of dephasing, which is expected to be particularly 
relevant in heterostructures as long as the motion of the
carriers can be described as ballistic. 
As for the interaction leading to
superconductivity we consider the simplest case of an instantaneous
local interaction, as described by an attractive Hubbard model.

While the effect of a voltage bias on a Bardeen-Cooper-Schrieffer
(BCS) superconductor is well established\cite{BardeenRMP,Tinkham96},
much less is known about intermediate and strong-coupling superconductors. These
regimes require to go beyond the BCS approximation and the use of
nonperturbative approaches like  Dynamical Mean-Field Theory
(DMFT)\cite{Georges1996RMP}.
This method allows to follow the evolution of statical and dynamical
properties of the superconducting state as a function of physical
parameters\cite{Garg2005PRB,Bauer2009PRB,Toschi2005NJoP,Toschi2005PRB}
and external stimuli.
The DMFT has been successfully applied to heterostructures formed by
normal metallic leads separated by a strongly correlated barrier
material in order to study transport and spectral properties at
equilibrium\cite{Freericks2004PRB,Freericks2006,Chen2007PRB,Okamoto2004PRB,Mazza2014}
and in a steady state\cite{Okamoto2008}. 
The effect of a voltage bias in a correlated metal has been
investigated in~\cite{Han2009PRB}, pointing out the existence of a critical
voltage for the metal-insulator transition. 
Recently a non-equilibrium extension of the DFMT
\cite{Freericks2006PRL,RevModPhys.86.779} has been used to
study the dynamics of driven strongly correlated
systems\cite{Joura2008PRL,Eckstein2010PRL,Amaricci2012PRB},
including inhomogeneous normal setup\cite{Eckstein2013PRB}. 

Our calculations show that an intermediate-coupling superconducting
state is essentially untouched by biases smaller than the energy
gap. Interestingly, as the voltage is increased, the phase transition
to a normal (non superconducting) state is preceded by the formation
of an intermediate {\it   bad}-superconducting state for biases
immediately larger than the energy gap. In this region, the bias leads
to an incoherent motion of the carriers which results in a
superconductor with a smaller energy gap and a finite scattering rate,
while the superfluid stiffness is slightly enhanced with respect to
equilibrium. Eventually the system turns into an insulating state of
incoherent preformed pairs for very large bias.
We characterize the two transitions via the evolution of the spectral
properties and of the characteristic energy scales contributing to the
stabilization of the bad superconductor and draw a phase diagram in
the interaction-voltage plane. 

In the following \secu{Sec2} we discuss the approximation
scheme, we derive the main model and discuss its method of solution
with the Nambu-Keldysh stationary DMFT. 
In \secu{Sec3} we present our main results, namely the progressive loss
of coherence of the quasi-particles leading to the breakdown of the
superconducting state. We discuss the formation of an intermediate
bad-superconducting state and the phase-diagram of the model for 
the intermediate-to-strong regime of attraction. 
In \secu{Sec4} we characterize the formation of the
bad-superconducting state by means of its energy
balance. Finally in \secu{Sec5} we present our conclusions and discuss
some future perspectives of this work.   

\section{Method.}\label{Sec2}
We consider a three-dimensional superconductor in presence of an
electric field closely following an approach proposed in
Ref. ~\onlinecite{Han2009PRB} for a correlated metallic system.
We work in the Coulomb gauge, where the constant field corresponds to
static potential difference between the external normal source/drain
leads.
As we are interested in the bulk properties of the system, we shall 
assume that the external leads are located far in space from the 
bulk region of interest.
The charge recombination takes place at the
interface of the superconductor with the normal leads. 

In the stationary limit the properties of such system are  determined
by the non-equilibrium electronic distribution $f(\omega)$, which in
turn can be separated a particle-hole symmetric component and an
asymmetric component describing the charge imbalance in the device. 
The spirit of this work is to focus on the bulk properties by neglecting
the effects of the charge-imbalance part. 

This simplification is justified by the fact that in the
superconducting regime the electrostatic potential at the
center of the device is essentially constant\cite{Keizer2006PRL},
while most of the potential drop occurs in proximity of the interface
to the normal leads. This makes the charge imbalance negligible in the
bulk. Note that this situation is different from the normal case,
for which the potential profile is almost linear in the
insulating regime and flat in the metallic regime\cite{Mazza2014}. 

Thus, in the stationary limit the effect of the voltage bias onto
the superconducting bulk can be described in terms of the existence of carriers at
energies $\pm\Phi/2$ of the leads, moving ballistically along the
direction set by the potential difference.  
Thus, independently of the actual dimensions of the system, we
classify the lattice electrons in two sets, namely of left- (L) and right-
(R) moving carriers with a local energy potential $V_L\!=\!+\Phi/2$
and $V_R\!=\!-\Phi/2$. This decomposition reflects the presence at the
bulk of a preferred direction corresponding to an electric field
associated to the bias. 
The physical destruction (creation) operators of an orbital electron at site $i$
with spin $\s$ can thus be decomposed in the superposition of L and R
electrons, \ie $c_{i\s}\! =\! \frac{1}{\sqrt{2}}\left( c_{i\s L}\! +\! 
c_{i\s R} \right)$.
We further assume an attractive interaction, which we model as on-site
attractive Hubbard $U$. The interaction obviously acts on the physical electrons
and gives rise to pairing and superconductivity. 
Our starting point is the Hamiltonian:
\eq{
\HH\! =\! -t\sum_{\bra ij\ket \s}\sum_{\a=L,R} \left(c^{+}_{i\s\a}c_{j\s\a} \!+\!
h.c.\right)
-|U|\sum_i n_{i\up} n_{i\dw}
}{Ham1}
where the first (kinetic) terms describes the ballistic motion of the
two species of movers, and the interaction is associated to the
densities of physical electrons $n_{i\s}\! =\! c^{\dagger}_{i\s} c_{i\s}$.
The model can be simplified by an orthogonal transformation to {\it
  even} (E) and {\it odd} (O) states  $c_{i\s E/O} \!=\!
\frac{1}{\sqrt{2}}\left(c_{i\s L} \!\pm\! c_{i\s  R}\right)$, where the E
combination coincides with the physical electron. 
In this basis, the interaction term obviously contains only
E electrons and the motion of the O component is decoupled and
trivial, as it is controlled only by the hopping term.
We consider the half-filling condition of one particle per site
corresponding to set the chemical potential to $\mu=|U|/2$.

Thanks to the bulk approximation scheme introduced above we can solve the model
(\ref{Ham1}) by means of the DMFT in the Keldysh formalism.
Within DMFT, the lattice problem is reduced to an effective
single impurity model\cite{Georges1996RMP,Toschi2005PRB}. 
DMFT allows to study directly the superconducting state of the system
introducing anomalous Green's function and self-energy associated to
the evolution of Cooper pairs. Then we can use a Nambu spinor
formalism, in which diagonal elements of the 2$\times$2 self-energy
matrix are normal components and off-diagonal ones are the
superconducting components\cite{Toschi2005PRB,Toschi2005NJoP}.
Starting the self-consistency equation with a finite anomalous Green's
function, the self-consistent loop will find superconducting solutions
if they are allowed by the physics, and it will correctly describe the
disappearance of superconductivity as a function of control
parameters. 

We use iterative perturbation theory method (IPT), extended to
deal with superconductivity\cite{Garg2005PRB}, to solve the effective
impurity problem.
This method provides a simple but reasonably accurate solver, which
is straightforwardly extended to the Keldysh formalism in real-time and
real-frequency and, importantly, gives direct access to the local
spectral functions. 
In the IPT method the self-energy function $\hat{\Sigma}$ is expressed as:
$
\hat{\Sigma}\! =\! \hat{\Sigma}_{\textsc{\tiny HFB}}\! +\! U^2\hat{\Sigma}^{(2)}
$, where $\Sigma^r_{\textsc{\tiny HFB}}\! =\! -\frac{|U|}{2}n\tau_3\! -\! \Delta \tau_1$
is the the Hartree-Fock-Bogoliubov (HFB) term, $\Delta\!=\!-|U|\bra
c_{o\up} c_{o\dw}\ket$  is the superfluid amplitude of the impurity and
$\tau_{i=1,2,3}$ are the Pauli matrices. 
$\hat{\Sigma}^{(2)}$ is the 2$^{nd}$-order contribution to the
self-energy, which in terms of the Keldysh components of the HFB-corrected
Weiss-Field: 
$
\hat{\tilde{\GG}}\!=\!
(\hat{\GG}^{-1}_0\!-\!\hat{\Sigma}_\mathrm{{\tiny HFB}})^{-1}
$
reads:
\eq{
\begin{split}
{\Sigma^{(2)}_{11}}^\lessgtr(t) &\!=\!\left[
\tilde{\GG}^\lessgtr_{11}(t)\tilde{\GG}^\lessgtr_{22}(t)  \!-\! 
\tilde{\FF}^\lessgtr_{12}(t)\tilde{\FF}^\lessgtr_{21}(t) 
\right]\tilde{\GG}^\gtrless_{22}(-t)\\
{S^{(2)}_{12}}^\lessgtr(t)      &\!=\! \left[
\tilde{\FF}^\lessgtr_{12}(t)\tilde{\FF}^\lessgtr_{21}(t) \!-\! 
\tilde{\GG}^\lessgtr_{11}(t)\tilde{\GG}^\lessgtr_{22}(t) 
\right]\tilde{\FF}^\gtrless_{12}(-t)
\end{split}
}{Sigma}
while the  retarded components are obtained from the general 
relation $F^r(t)\! =\! \theta(t)(F^>(t)\! -\! F^<(t))$.

The self-consistency condition is fulfilled by imposing to 
the retarded component of the Weiss field to obey the relation: 
$\hat{\GG}_0^{r-1}(\omega)\!=\!{\hat{G}}^{r-1}(\omega)\!+\!\hat{\Sigma}^r(\omega)$,
where $\hat{G}^r(\omega)\!=\!\int\! d\e D(\e) \hat{G}^r(\e,\omega)$ is the
real-frequency local retarded Green's function and $D(\e)$ is the
non-interacting density of state of the lattice.
Here we use a semicircular density of states $D(\e)=2\sqrt{D^2\! -\!
  \e^2}/\pi D^2$ of half-bandwidth $D$ (that we set as our energy
unit), which has been shown to describe accurately three-dimensional
systems, and substantially simplifies the self-consistency equations,
thereby reducing numerical errors which can be particularly dangerous
out of equilibrium. Because of its local nature, the DMFT calculations 
depend very weakly on  the lattice structure. Thus, our results
are qualitatively valid for any other choice of the underlying lattice structure.

Finally, the other Keldysh components of
the Weiss field are obtained through the spectral representation as:
\eq{
\left[\begin{array}{c}
\tilde{\GG}_{11}^{\lessgtr}(t) \\ 
\tilde{\FF}_{12}^{\lessgtr}(t)
\end{array}\right]
\!=\! \mp \frac{i}{2\pi}\!\int \!d\omega \!
\left[\!
\begin{array}{c}
\Im{\tilde{\GG}^r}(\omega) \\ \Im{\tilde{\FF}^r}(\omega)
\end{array}
\!\right]\!
f(\pm\omega)
e^{-i\omega t}
}{G0eq}
where $f(x)=\sum_{\a=\pm}f_{F\!D}(x\!+\!\a\Phi/2)$ is the
effective-bath distribution function, given by the symmetric combination of the Left/Right moving
electrons distribution, as follow from the initial assumptions, and
$f_{F\!D}$ is the Fermi-Dirac function. 
This equation expresses, within our local
approximation of the lattice problem, the effects of the continuous
energy mismatch brought in by the presence of the voltage bias. 
A similar expression can be derived for the order parameter within the BCS
theory\cite{BardeenRMP,Tinkham96}, to describe the progressive
destruction of the superconducting order as effect of a voltage
bias. Our \equ{G0eq} reduces to an approximated version of such BCS
equation for the order parameter in the limit of very weak attraction,
in agreement with the nature of the bulk approximation.

Finally, we determine the remaining Keldysh components of the normal
and anomalous Green's function using the following relations: 
\eq{
\begin{split}
&G_{22}^\lessgtr(\omega)  = -G_{11}^\gtrless(-\omega)\\
&F_{21}^\lessgtr(\omega)  = -{F_{12}^\lessgtr}^*(\omega)\\
&G_{22}^r(\omega) =  {G_{11}^r}^* (-\omega)\\
&F_{21}^r(\omega) =  {F_{12}^r}^*(-\omega)
\end{split}
}{GFrelations}

\begin{figure}
  \includegraphics[width=0.48\textwidth]{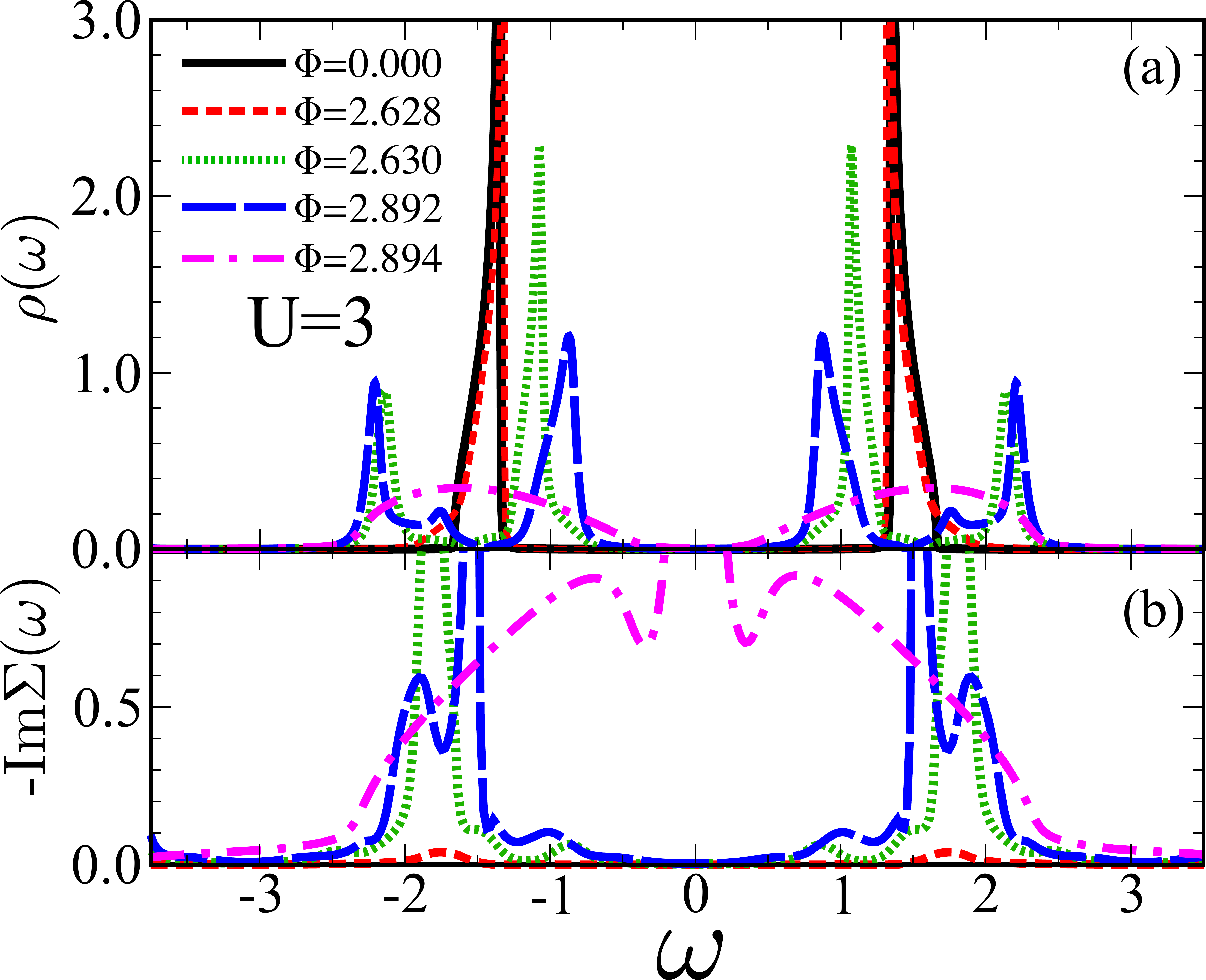}
  \caption{(Color online) Evolution of the spectral function
    $\rho(\omega)\!=\!-\Im{G^r_{11}(\omega)}/\pi$ (a) and of the imaginary
    part of the normal self-energy $-\Im{\Sigma}^r_{11}(\omega)$ (b) as a function of the
    voltage bias $\Phi$ for $U\!=\!3$. 
  }
  \label{fig1}
\end{figure}

\section{Results.}\label{Sec3}
In the absence of any bias ($\Phi\!=\!0$) the attractive Hubbard model describes a 
superconducting (SC) to normal phase-transition at a critical temperature 
$T_c(U)$\cite{Toschi2005PRB,Toschi2005NJoP} for any value of the
carrier density and the interaction. 
While the modulus of the order parameter $P\!=\! 1/N\sum_i \langle
c_{i\uparrow}c_{i\downarrow}\rangle$ and the energy gap
$E^0_g$ monotonically increase as a function of $U$, the
critical temperature displays a maximum at an intermediate value of
$U$ of the order of the bandwidth, roughly separating the BCS-like behavior at 
weaker coupling from the strong-coupling or Bose-Einstein-Condensate
(BEC) regime. In this latter case the large pairing strength locks the fermions in
strongly bounded local pairs, which delocalize with a suppressed hopping
amplitude proportional to $t^2/U$. 
  
In this work we focus on the intermediate-to-strong coupling regime,
where the BCS approach does not apply. 
Because of the assumptions behind the bulk approximation scheme, we
expect our method to better describe the intermediate-to-strong
attracting regime, where the coherence length scale is shorter. 

We investigate how the presence of an
external bias destabilizes the SC state for different
values of $U$ in the half-filled case where the number of electrons is
identical to the number of sites
\footnote{For the attractive model this regime has an extra symmetry as the
SC state is degenerate with a commensurate charge-density wave, but,
as far as the SC solution is concerned, the results are 
representative of any density.}.

In \figu{fig1} we show the evolution of the single-particle spectral
density $\rho(\omega)\!=\!-\Im{G^r_{11}(\omega)}/\pi$ and of the
single-particle self-energy
$\Im{\Sigma^r_{11}(\omega)}$ by increasing the bias at $U\!=\!3$.
At zero bias $\Phi\!=\!0$ the spectrum displays narrow coherence peaks
at the equilibrium gap edges $\pm E^0_g/2$. 
Within IPT the incoherent part of
the spectral weight is split
into a narrow peak and a satellite at higher energy (not shown in the
figure). This is a consequence of the vanishing of the imaginary part
of the normal self-energy for $|\omega|\!<\!\frac{3E^0_g}{2}$ as follows
from the Eq.~\ref{Sigma} for a gapped system\cite{Bauer2009PRB}.

\begin{figure}
\includegraphics[width=0.5\textwidth]{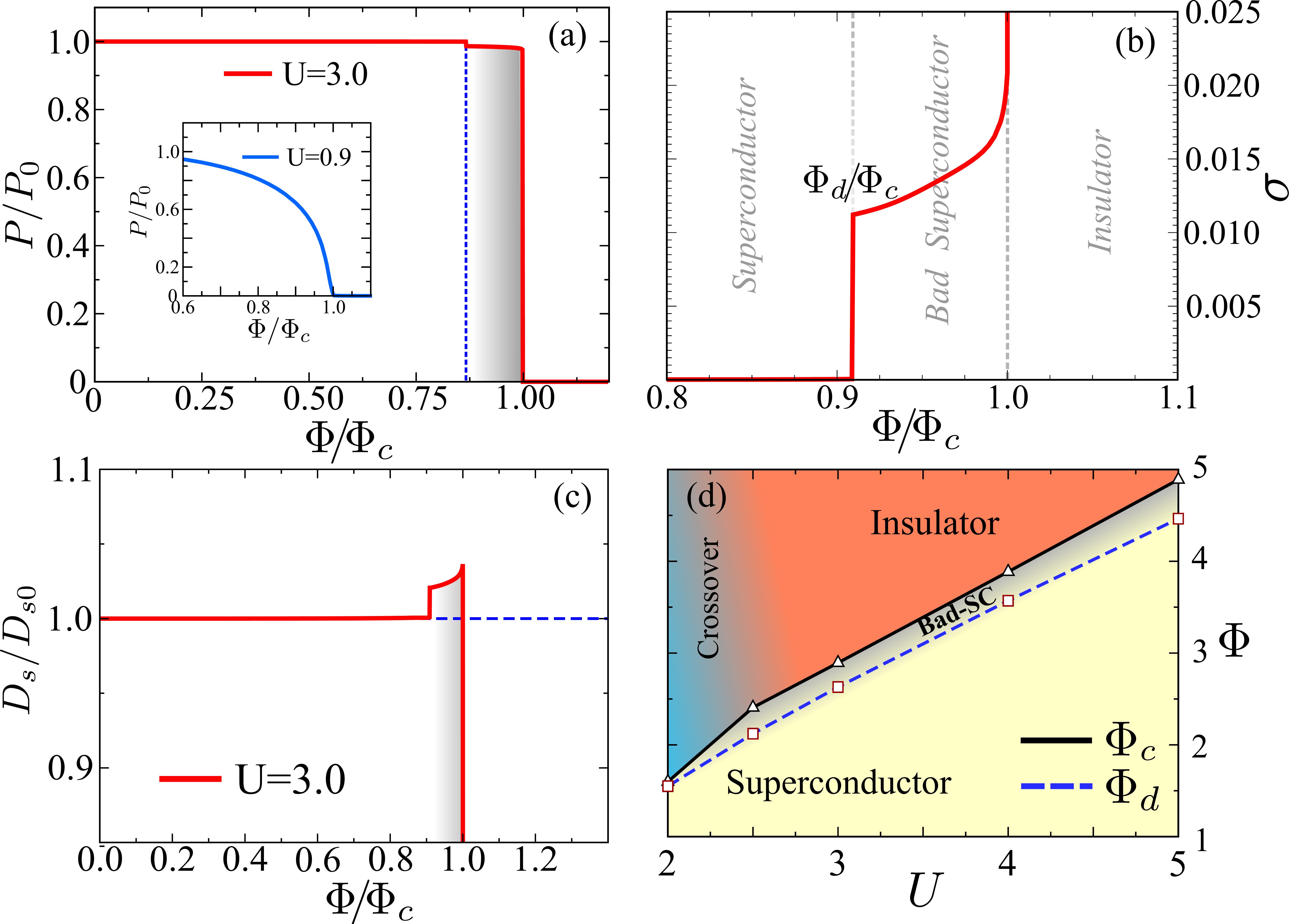}
\caption{(Color online) 
  (a) Order parameter $P$, normalized to its 
  equilibrium value $P_0$, for  $U\!=\!3$ ($P_0\!=\!1.414$) and
  $U\!=\!0.9$ (inset, $P_0\!=\!0.169$). 
  (b) Scattering rate $\sigma$ for $U\!=\!3$.
  (c) Normalized superfluid stiffness $D_s/D_0$ for $U\!=\!3$.
  (d) $\Phi$-$U$ phase-diagram of the model. 
  Dashed line shows the transition to a bad superconducting regime. 
  Solid line indicates the transition to the normal phase.}
\label{fig2}
\end{figure}
The effect of the bias on our strong coupling SC is extremely small as
long as the bias $\Phi$ is substantially smaller than the equilibrium
gap  $\!E^0_g\! \simeq\! 2.894$ for our choice of parameters. 
In this regime the spectral function essentially coincides with the equilibrium solution. 
In the same process a finite imaginary part of the self-energy very
slowly develops at low-energy near $\pm E^0_g/2$ (see
Fig.~\ref{fig1}(b)), associated to an increased scattering rate
$\sigma\!=\!-\Im{\Sigma(\omega\!=\!E^\Phi_g)}$ 
, \ie to a reduced lifetime of the single particle excitations at the
gap edge. 
This effect is however negligible as long as the bias is smaller than a critical value 
$\Phi_d\!=\! 2.630 < E^0_g$ at which the system undergoes a
discontinuous transition where the gap edge sharply drops to a lower
value (green dotted curve in \figu{fig1}). 
At the same time also other observables (see panels (a)-(c) in
\figu{fig2}) have a discontinuous change: 
The order parameter $P$ has a small drop, while
the scattering rate $\sigma$ jumps to a larger value  as
well as the superfluid stiffness $D_s$.
In this regime, the weight of $\Im\Sigma_{11}^r(\omega)$ in the range
$|\omega|\!\geq\!{E^0_g}/{2}$ becomes more pronounced. This corresponds to further
broadening of the coherence peaks in the spectral functions, \ie to a progressive
shortening of quasi-particles life-time.

Further increasing the voltage bias we reach a (second) critical value
$\Phi_c\! >\! \Phi_d$ for which the system is no longer able
to sustain the SC state, and undergoes a superconductor-to-normal phase transition,  
characterized by a profound change in the spectral functions and the observables. 
For our intermediate- and strong values of the interaction $U$, the final normal state is indeed
an insulator formed by incoherent localized pairs. For $\Phi\!>\!\Phi_c$ the spectral
density is characterized by the onset of completely incoherent 
bands centered at $\omega\!=\!\pm
U/2$\cite{Keller2001PRL,Capone2002PRL}, 
analogous to the Hubbard bands in the repulsive Hubbard model. The
self-energy
accordingly diverges for $\omega\to 0$, see Fig.~\ref{fig1}(b). Here
the superfluid order parameter $P$ and the superfluid stiffness drop
to zero, while the scattering rate diverges, signaling the incoherent
nature of this insulator.

In the inset of \figu{fig2}(a) we show DMFT results for the order
parameter obtained for a smaller value of $U\!=\!0.9$, for which the BCS
scheme can be qualitatively used. Indeed our approach qualitatively reproduces the
correct continuous reduction of the order parameter, as predicted by the BCS
theory\cite{BardeenRMP}, confirming that the results obtained at
larger $U$ in DMFT do not depend on the bulk approximation. 
This result is directly linked to the structure of the equation
\equ{G0eq}, which in the weak-attraction regime reduces to an
approximated result within the BCS theory\cite{BardeenRMP,Tinkham96}. 

In panel (d) of Fig. \ref{fig2} we report a phase diagram as a
function of the attraction U and the bias. The same transitions we
described are indeed present for any $U\!>\!2$. Two transitions lines
$\Phi_d(U)$, $\Phi_c(U)$ define a standard superconducting region, a
bad superconductor and a poorly conducting or insulating normal
state. Indeed the existence of the bad superconducting intermediate
state  appears to be limited to parameters for which the normal state
is not a normal metal, but an incoherent insulator of preformed
pairs\cite{Capone2002PRL}. 

\section{Energy balance.}\label{Sec4}
\begin{figure}
\includegraphics[width=0.4\textwidth]{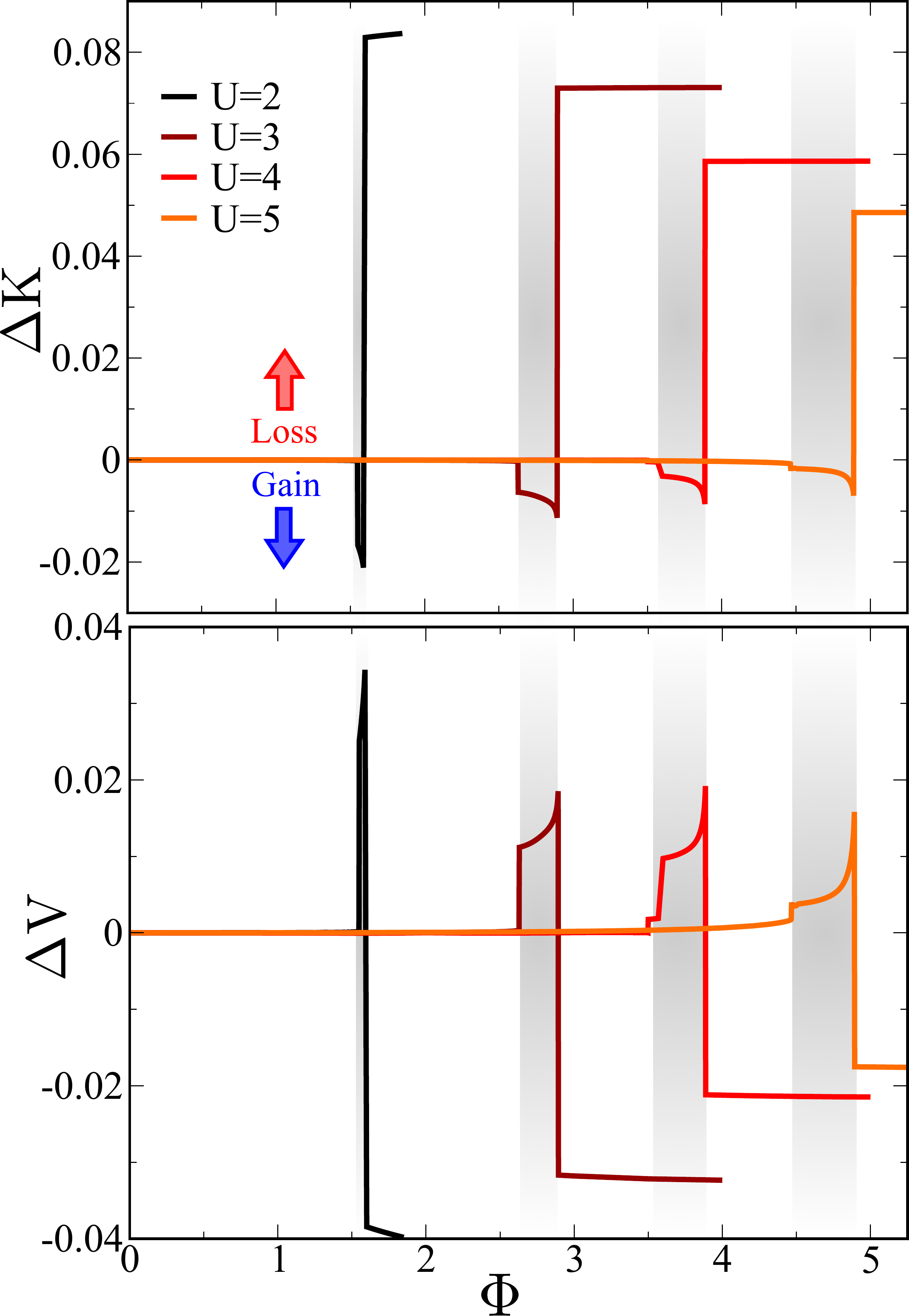}
\caption{(Color online) Kinetic (left) and potential (right) energy
  variation, respectively $\Delta K$ and $\Delta V$, as a function of the voltage
  bias $\Phi$. 
}
\label{fig3}
\end{figure}

We now characterize the bad superconducting state and the reason
behind its stabilization.
In \figu{fig3} we show the variations of the
kinetic and potential energy with respect to equilibrium as a function 
of the bias, $\Delta E\!=\!E(\Phi)\!-\!E(0)$. 
Interestingly, our results show that the bad superconductor has a
lower kinetic energy and a higher potential
energy with respect to its equilibrium counterpart. The increased
potential energy amounts to a reduction of the energy gain associated
to Cooper pair formation, and it is consistent
with the reduction of the SC order parameter and of the
spectral gap that we discussed before. However, the system is
able to gain kinetic energy in the same process, compensating the
dephasing effect of the bias.

This latter fact is crucial for the existence of the bad
superconductor and is intimately related to the intermediate- and
strong-coupling nature of the superconductor. In this regime, in fact,
the system can be described as a collection of bound pairs which
manage to establish phase coherence despite their limited
mobility. This is in sharp contrast with the BCS regime, where
delocalized pairs move across the system.

The external bias can help the preformed pairs to move, even if incoherently.
This indeed leads to the gain in kinetic energy that we have reported
in \figu{fig3} for the bad superconductor.
The increased incoherent motion contrasts with the binding in pairs
and leads to a loss in the potential energy. As a matter of fact the biased
superconductor behaves as an equilibrium system with a smaller
effective coupling, which indeed corresponds to a smaller potential
energy gain, but also to a gain in kinetic energy.
However, the bias also leads to processes giving rise
to a finite scattering rate (see \figu{fig2}(b)) which
testify the incoherent nature of this intermediate superconducting
state. It is precisely this reduced coherence which led us to define
this state as a ``bad superconductor''.

Interestingly, the bad superconductor turns out to have a
{\it{larger}} superfluid stiffness $D_s$ than the equilibrium system,
as reported in \figu{fig2}(c). This result can be rationalized in
terms of the two effects we described above. On one hand the bad
superconductor is expected to have a reduction of superfluid stiffness
as a consequence of the incoherence measured by the scattering
rate. On the other hand, the energetic balance suggests that, to some
extent, the bad superconductor can be regarded as an effectively
weaker-coupling system. Indeed $D_s$  naturally {\it{increases}}
reducing the coupling\cite{Toschi2005PRB}. The two effects are
therefore competitive, but our DMFT calculations show that the
effective reduction of the coupling prevails, leading to an overall
increase of the superfluid density. We have verified that the observed
increase is much smaller than the one associated to the effective
reduction of the coupling.

Upon further increasing the voltage bias, the
delicate balance which stabilizes the bad superconductor can no longer
be sustained and the system undergoes a phase-transition in which all
the residual kinetic energy in transformed into potential energy
leading to formation of an insulator of preformed pairs. 

\section{Conclusions.}\label{Sec5}
In this work we have investigated a model for non-equilibrium
superconductivity. We solved the model using dynamical mean-field
theory within iterative perturbation theory calculations. 
We find that the non-equilibrium stationary superconducting state is
only destabilized by a voltage bias of the order of the gap. Our
calculations show that the transition to the normal phase takes place
in two steps: crossing a first critical line the system enters in a
bad superconducting regime, characterized by more incoherent
quasi-particles excitations and smaller superfluid
amplitude. Eventually the bad superconductor turns into a insulator of
completely incoherent preformed paris. The stabilization of this state
is shown to intimately related to the intermediate/strong-coupling
nature of pairing, in which tightly bound pairs are formed already in
the normal state.

\paragraph*{Acknowledgments.} 
A.A. acknowledges insightful discussions with J.~Han.
This work is supported by the European Union under
FP7 ERC Starting Grant n.240524 ``SUPERBAD" and by
FP7 GO FAST, grant agreement no. 280555.
\bibliography{localbib}
\end{document}